\documentclass{article}

\RequirePackage{hyperref}
\usepackage[utf8]{inputenc} 
\usepackage{url,graphicx,amsmath,fullpage}
\usepackage{subfig}
\usepackage{hyperref}
\usepackage{comment}

\begin{document}




\title{MaxSSmap: A GPU program for mapping divergent short reads
to genomes with the maximum scoring subsequence}
\author{Turki Turki$^{\footnote{Computer Science Department, King Abdulaziz University,
P.O. Box 80221, Jeddah 21589, Saudi Arabia}}$\and
Usman Roshan$^{\footnote{To whom correspondence should be addressed; usman@cs.njit.edu,
Ph: 862-485-0590, Fax: 973-596-5777}}$ 
$^{\footnote{Department of Computer Science, New Jersey Institute of Technology, Newark, NJ 07102}}$
}

\date{}
\maketitle




\begin{abstract} 
Programs based on hash tables and Burrows-Wheeler are very fast for mapping
short reads to genomes but have low accuracy in the presence of mismatches
and gaps. Such reads can be aligned accurately with 
the Smith-Waterman algorithm but it can take
hours and days to map millions of reads even for bacteria genomes. 
We introduce a GPU program called MaxSSmap with the aim of achieving
comparable accuracy to Smith-Waterman but with faster runtimes. Similar to
most programs MaxSSmap identifies a local region of the genome 
followed by exact alignment. Instead of using hash tables or Burrows-Wheeler
in the first part, MaxSSmap calculates maximum scoring subsequence 
score between the read and disjoint fragments of 
the genome in parallel on a GPU and selects
the highest scoring fragment for exact alignment. We evaluate MaxSSmap's
accuracy and runtime when mapping simulated Illumina {\it E.coli} and human 
chromosome one reads of different lengths and 10\% to 30\% mismatches 
with gaps to the {\it E.coli} genome and human chromosome one. 
We also demonstrate applications on real data by 
mapping ancient horse DNA reads to modern genomes
and unmapped paired reads from NA12878 in 1000 genomes.
We show that MaxSSmap attains comparable high accuracy and low error to
fast Smith-Waterman programs yet has much lower runtimes. We show
that MaxSSmap can map reads rejected by BWA and NextGenMap with high accuracy
and low error much faster than if Smith-Waterman were used. On short read
lengths of 36 and 51  both MaxSSmap and Smith-Waterman have lower accuracy
compared to at higher lengths. On real data MaxSSmap produces many
alignments with high score and mapping quality that are not given by NextGenMap
and BWA. The MaxSSmap source code is freely available from 
\url{http://www.cs.njit.edu/usman/MaxSSmap}.
\end{abstract}


%



\section{Introduction}
In next generation sequencing experiments we may encounter divergent reads in
various scenarios. These include structural variation studies, comparison of 
distantly related genomes, absence of same species reference genome, 
sequence error in long reads, genome variation
within same species, ancient DNA mapping, and mRNA-seq experiments
\cite{stampy,rnaseq,reynoso,collins,seabury,liang,longreaderror,genomevariation,
ancientDNA,zamin}. Programs \cite{fonseca,ayat} based on hash tables and 
Burrows-Wheeler transform are very fast 
but have low accuracy on such reads that tend to contain many mismatches and
gaps \cite{stampy,nextgenmap}. The Smith-Waterman algorithm 
\cite{smithwaterman} can map divergent
reads accurately but is considerably expensive. Even high performance multi-core
and Graphics Processing Unit (GPU) implementations
can take hours and days to align millions of reads even to bacteria genomes. 
As a solution we introduce a GPU program called MaxSSmap with the aim of achieving
comparable accuracy to Smith-Waterman on divergent reads but with faster runtimes.


We divide the genome into same size disjoint fragments and then map 
a read to all fragments in parallel on a GPU with the maximum
scoring subsequence score \cite{bentley86,batesandconstable85}. 
A GPU can run several hundred threads at the same time and allows for massive 
parallelism in computer programs (see \url{http://www.gpucomputing.net}).
The maximum scoring subsequence is roughly the same as Smith-Waterman
except that it does not consider gaps.
Once we identify the first and second highest scoring fragments 
--- we need the second to eliminate repeats --- we perform 
Needleman-Wunsch alignment of the read to the identified region of genome.
We present a GPU program called MaxSSmap that implements 
this idea along with several heuristics and shortcuts that lead to faster runtimes 
without sacrificing accuracy. 

On reads with fewer than 10\% mismatches our program offers no advantage over
hash-table approaches. Programs like NextGenMap that use hash-tables
in their first phase can map such reads very quickly with high 
accuracy compared to other leading programs \cite{nextgenmap}. 
Thus we focus on reads with divergence between
10\% and 30\% as well as gaps of lengths up to 30 both in the read and genome. 

We compare MaxSSmap to two fast Smith-Waterman programs. The first is the 
recently published Smith-Waterman library for short read
mapping and protein database search called SSW \cite{ssw}. 
This uses a fast Single-Instruction-Multiple-Data Smith-Waterman algorithm to align a given 
read to the entire genome. The authors of the program demonstrate improved and comparable 
runtimes to state of the art fast Smith-Waterman programs for 
mapping DNA sequences to a genome. In addition this produces output in SAM format and 
has also been applied it to real data in the context of realigning unmapped reads \cite{ssw}. 
The second is a fast GPU Smith-Waterman program for protein sequence database search called 
CUDA-SW++ \cite{cudasw++}. We note that this is not 
designed for mapping DNA sequence reads. However, we adapt it to short read mapping 
by considering fragments of the genome as database
sequences and read as the query. 

Exact Smith-Waterman methods take much longer than hash-table 
and Burrows-Wheeler based programs to align millions of reads to genome 
sequences. In this setting we study several meta-methods that first align reads with a 
fast program and then map rejected ones with a slower but more accurate one such as 
MaxSSmap and SSW.

We study accuracy and runtime for mapping simulated Illumina {\it E.coli} and human reads of
various lengths to the {\it E.coli} and human chromosome one. Our focus is on reads 
with 10\% to 30\% mismatches and gaps up to length 30.
We show that MaxSSmap attains comparable high accuracy and low error as CUDA-SW++
and SSW but is several fold faster than the two programs respectively. We show 
that MaxSSmap can map reads rejected by NextGenMap \cite{nextgenmap} 
with high accuracy and low error and much faster than if Smith-Waterman were used.
We also study MaxSSmap on various read lengths and demonstrate applications 
on real data by mapping ancient horse DNA reads to
modern genomes and unpaired mapped reads in 1000 genomes subject NA12878.

Below we provide basic background and describe our program in detail. We then present 
our experimental results on simulated and real data.

\section{Methods}

\subsection{Background}
Before we describe MaxSSmap we provide background on the maximum 
scoring subsequence and GPUs and CUDA.

\subsubsection{Maximum scoring subsequence}
The maximum scoring subsequence for a sequence of 
real numbers $\{x_1,x_2,...,x_n\}$ is defined to be
the contiguous subsequence $\{x_i,...,x_j\}$ that maximizes the sum $x_i+...+x_j$ 
($0\leq i,j \leq n$). A simple linear time approach will find the maximum scoring
subsequence \cite{bentley86,batesandconstable85}. To apply this to DNA sequences
consider two of the same length aligned to each other without gaps. Each aligned
character corresponds to a substitution whose cost can be obtained from 
a position specific scoring matrix that accounts for base call probabilities,
or a substitution scoring matrix, or a trivial match or mismatch cost. The maximum
scoring subsequence between the two DNA sequences can now be obtained through this
sequence of substitution scores \cite{bentley86,batesandconstable85}. 

\subsubsection{Graphics Processing Units (GPUs) and CUDA}
CUDA is a programming language that is developed by NVIDIA. 
It is mainly C with extensions for programming on NVIDIA GPUs. 
We use CUDA version 4.2 for all GPU programs in this study.
The GPU is designed for running in parallel hundreds of short functions called threads.
Threads are organized into blocks which in turn are organized into grids. 
We use one grid and automatically set the number of blocks to the total number
of genome fragments divided by the number of threads to run in a block. The
number of threads in a block can be specified by the user and otherwise we set it to 
256 by default. 

The GPU memory is of several types each with different size and access times:
global, local, constant, shared, and texture. Global memory is the largest and can be as 
much as 6GB for Tesla GPUs. Local memory is the same as global memory but limited to
a thread. Access times for global and local memory are much higher than the
those for a CPU program to access RAM. However, this time can be
considerably reduced with coalescent memory access that we explain below.
Constant and texture are cached global memory and accessible
by any thread in the program. Shared is on-chip making it the fastest and is limited to 
threads in a block. 
More details about CUDA and the GPU 
architecture can be found in the NVIDIA online documentation \cite{cudaguide} and 
recent books \cite{cudabyexample,pmpp}.

\subsection{MaxSSmap algorithm}

\subsubsection{Overview}
Our program, that we call MaxSSmap, follows the same two part 
approach of existing mappers: first identify a local region of the
genome and then align the read with Needleman-Wunsch (or Smith-Waterman)
to the identified region. The second part is the same as current methods
but in the first part we use the maximum scoring subsequence as
described below.

MaxSSmap divides the genome into fragments of a fixed size given by
the user. It uses one grid and automatically sets the number of blocks to the total number
of genome fragments divided by the number of threads to run in a block. The
number of threads in a block can be specified by the user and otherwise we set it to 
256 by default. 

\begin{figure}[h]
\centering
\includegraphics[scale=.5]{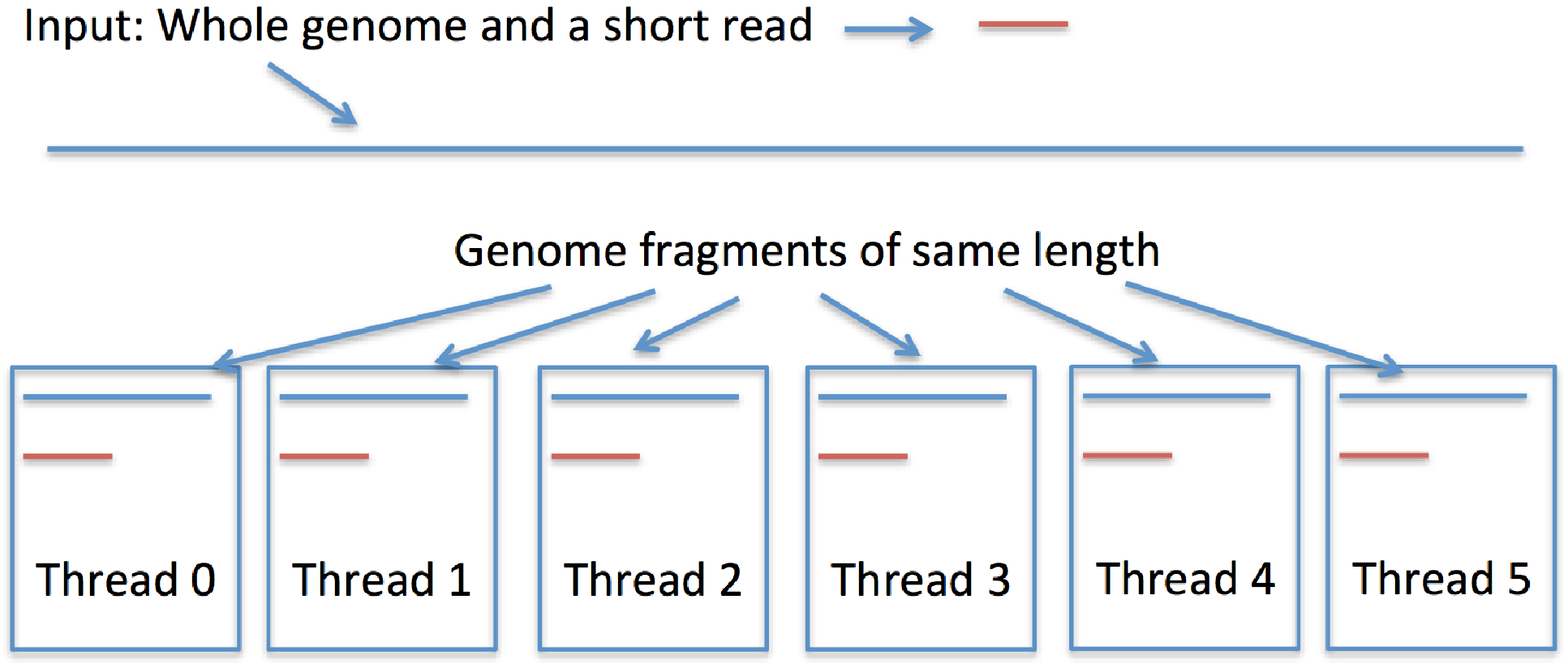}
\caption{Overview of the MaxSSmap program. In this figure the genome is divided into six
fragments which means six threads will run on the GPU. Thread with ID 0 maps the read
to fragment 0, slides it across fragment 0, and stops when it has 
covered all of fragment 0. We account for junctions between 
fragments and ensure that the read is fully mapped to the genome. 
\label{maxssmap2}
}
\end{figure}

\paragraph{First phase of MaxSSmap}
In Figure~\ref{maxssmap2} we show an overview of the MaxSSmap program.
Each thread of the GPU computes the maximum scoring 
subsequence \cite{bentley86} of the read and a unique fragment 
with a sliding window approach. In order to map across junctions 
between fragments each thread also considers neighboring 
fragments when mapping the read. 
When done it outputs the fragment number with the highest and second highest
score and considers the read to be mapped if the ratio of the second best to
best score is below .9 (chosen by empirical performance). 
This reduces false positives due to repeats. We later define a mapping quality score
that is based on this ratio. 

\paragraph{Second phase of MaxSSmap}
After the fragment number is identified we consider the region of the genome starting
from the identified fragment and spanning fragments to the right until we have
enough nucleotides as the read sequence. In the second part we align the
read sequence with Needleman-Wunsch to the genome region from the
first part. The default settings for match, mismatch, and gap costs that we also use
in this study are set to 5, -4, and -26. 

\paragraph{Incorporating base qualities and position specific scoring matrix}
We also consider the base qualities of reads in both phases of the program.
This can be done easily by creating a position specific scoring matrix for each read that also
allows for fast access using table lookup \cite{cudasw++}. For example 
let $x$ be the probability that the base at position $i$ is correctly sequenced. This can be
calculated by the phred score \cite{phred} that is provided with the reads. 
The score of a match against the nucleotide at position $i$ is 
$match\times x$ and mismatch is $mismatch\times \frac{x}{3}$. 

\paragraph{Mapping qualities, read lengths, SAM output, and source code}
MaxSSmap outputs in SAM format that is widely used for mapping DNA reads
to genomes \cite{sam}. In the MAPQ field of SAM \cite{sam} we use the
formula $-100\log_{2} p$ where $p$ is the probability of 
alignment being incorrect. We define this to be the ratio of
the scores of the second highest and top scoring fragments. For 
MaxSSmap we consider the read mapped only if mapping quality is above 
$-100\log_{2} .9=15.2$. MaxSSmap can also map reads of various lengths present in one fastq file. 
There is no need to specify the read length. However, the maximum read
length is limited to 2432 base pairs (bp) in the current implementation (see
paragraph on shared memory below). 
The source code is freely available from 
\url{http://www.cs.njit.edu/usman/MaxSSmap}.
\\
\\
We implement several novel heuristics and take advantage of the GPU architecture
to speed up our program which we describe below. 

\subsubsection{GPU specific heuristics}
\begin{figure}[h]
\centering
\includegraphics[scale=.5]{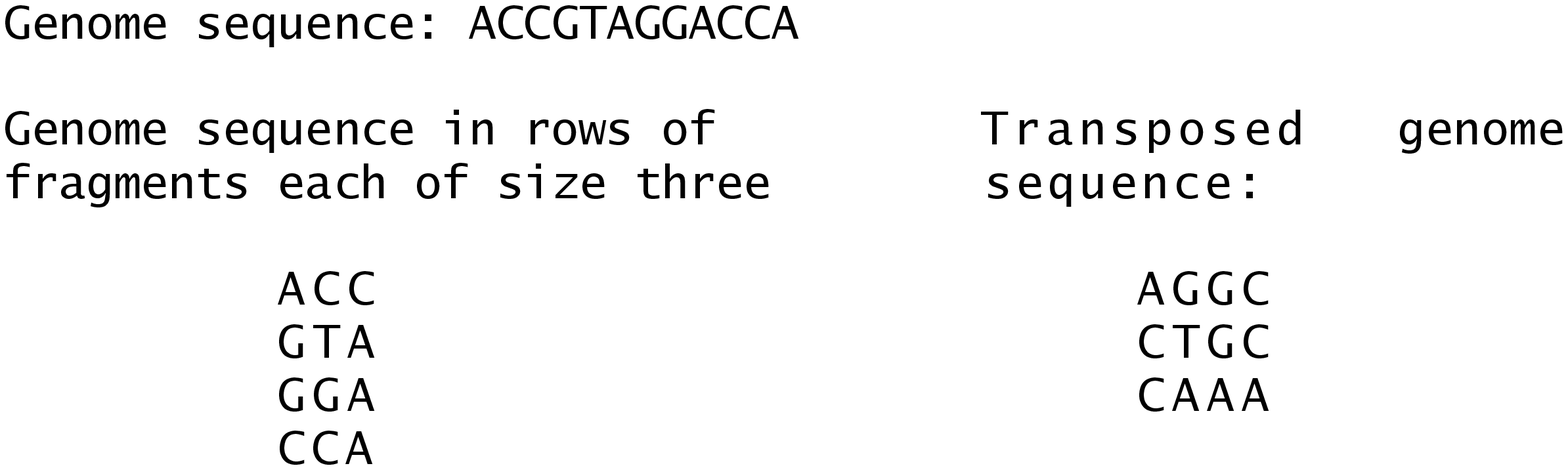}
\caption{Genome sequence in transpose format to enable coalescent memory access.
In MaxSSmap threads with IDs 0 through 3 would at the same time read characters
A, G, G, and C of the transposed genome to compare against the read. Since the four
characters are in consecutive memory locations and so are the thread IDs, our program 
makes just one read from global memory instead of four separate ones.}
\label{coalescentfigure}
\end{figure}

\paragraph{Coalescent global memory access \label{coalescent}}
Coalesced memory access is a key performance consideration when programming
in CUDA (see the CUDA C Best Practices Guide \cite{cudabestpractices}). 
Roughly speaking, each thread of a GPU has its own unique identifier 
that we call $thread\_id$. In order to have coalescent memory 
access our program must have threads with
consecutive identifiers access consecutive locations in memory (roughly speaking). 
We achieve this by first considering the genome sequence as rows of fragments
of a fixed size. We then transpose this matrix to yield a transposed genome sequence
that allows coalescent memory access. The transposed genome is transferred 
just once in the beginning of the program from
CPU RAM to GPU global memory. It has negligible overhead time compared to 
the total one for mapping thousands of reads.
See Figure~\ref{coalescentfigure} for a toy genome ACCGTAGGACCA 
and fragment length of three. If the genome is not a multiple of the fragment
length we pad the last fragment with N's. Our CUDA program runs a total of
$numfragments$ threads. In the example shown in Figure~\ref{coalescentfigure} 
there are four fragments. Thus our CUDA program would run four 
threads simultaneously with identifiers zero through three.
Each thread would access the transposed genome sequence first at location $thread\_id$,
then at $thread\_id + numfragments$, followed by location 
$thread\_id + 2numfragments$, and so on. 

\paragraph{Byte packing for faster global memory access}
In the GPU we store the genome sequence in a single array of int4 type instead of char.
This leads to fewer global memory accesses and thus faster runtimes. 
To enable this we append `N' characters 
onto the genome and query until both lengths are multiples of 16. This also requires
that the fragment length be a multiple of 16.

\paragraph{Look ahead strategy to reduce global memory penalties}
As mentioned earlier MaxSSmap uses a sliding window approach from left to right 
to map a read to a given fragment on the genome. In its implementation we compute the score
of the read in the current window and sixteen windows to the right at the same time. 
Therefore instead of shifting the window by one nucleotide we shift it by sixteen.
This leads to fewer global memory calls and also allows us to unroll loops.
See file MaxSSMap\_shared\_int4\_fast.cu in the source code for exact implementation.

\paragraph{Shared memory}
We store the query in shared memory to allow fast access. As mentioned earlier the GPU
access time to shared memory is fastest. This, however, imposes a limitation on the read length 
because shared memory size is much smaller than global memory. The 
Fermi Tesla M2050 GPUs that we use in this study have a maximum of 49152 bytes 
shared memory per block. The data structure stores the query in a profile format and 
so occupies a total of  $(readlength +16)\times 4 \times 5$  bytes. The 4 accounts for number
of bytes in a float, 5 is for bases A, C, G, T, and N, and 16 is for additional space
used by the look-ahead strategy and to eliminate if-statements in the code. 
Thus the maximum allowable DNA read length of the current 
implementation is 2432 bp (largest multiple of 16 below the cap size of 2441 bp). 
The query length can be increased at the expense of
running time by storing the query in constant memory, which is of size 65536 byes, 
or in global memory.

\subsubsection{Parallel multi-threaded CPU implementation of MaxSSmap}
We have also implemented a parallel multi-threaded CPU implementation of MaxSSmap
with the OpenMP library \cite{openmp08} (OpenMP available from 
\url{http://www.openmp.org}). Each thread maps the given read to a unique 
fragment of the genome. The  number of threads is automatically set to the genome
size divided by the specified fragment length. Thus if the fragment length is
4800 then for {\it E.coli} (approximately 5 million bp) it runs about 1042 threads
on the available CPU cores. This also uses the look ahead strategy
as described above. However, the coalescent and shared memory 
techniques don't apply to this version since they are specific to a GPU.

\subsection{Programs compared and their versions and parameters \label{parameters}}
The literature contains many short read alignment programs that have been
benchmarked extensively \cite{fonseca,ayat}. Instead of considering many different
programs we select the widely used program BWA \cite{bwa} that uses the 
Burrows-Wheeler transform. We also select NextGenMap that uses hash-tables and is shown
to be accurate on reads upto 10\% mismatches compared to other leading programs 
\cite{nextgenmap}. We use the multi-threaded version of BWA and enable the GPU option 
in NextGenMap. 

Other GPU programs for mapping short reads \cite{barracuda,mummergpu,cushaw,soapgpu}
are implementations of CPU counterparts designed for speedup and achieve the same accuracy.
Since they offer no improvement in accuracy they would perform poorly on divergent
reads. Furthermore, the CPU program runtimes are already in seconds vs. minutes and hours for
exact methods (such as ours and Smith-Waterman) and so we exclude these programs 
from the paper. 

From the category of exact mapping programs
we use SSW \cite{ssw} that uses a fast Single-Instruction-Multiple-Data (SIMD)
Smith-Waterman algorithm to align a given read to the entire genome and the
fast GPU Smith-Waterman program CUDA-SW++ \cite{cudasw++}.
As noted earlier this is designed for protein sequence database search and not
for aligning to large genome sequences. However, we adapt it to short read mapping 
by considering fragments of the genome as database
sequences and read as the query. 

Below we describe program parameters and how we optimized them where applicable. 
The exact command line of each program is given in the Online Supplementary Material
at \url{http://www.cs.njit.edu/usman/MaxSSmap}.

\paragraph{MaxSSmap}
For MaxSSmap we consider fragment lengths of 48 for {\it E.coli} genome, 480 for human
chromosome one, 4800 for horse and whole human genomes, and 
match and mismatch costs of 5 and -4 respectively. In the exact alignment
phase where we perform Needleman-Wunsch we consider the same match and mismatch
cost and a gap cost of -26. We selected fragment lengths to optimize runtime. We 
considered sizes of 16, 32, 48, 64, and 80 for the {\it E.coli} genome, 
lengths of 160, 240, 320, 400, and 480 for human chromosome
one, and lengths of  2400, 3600, 4800, 6000, and 7200 for the horse genome. 
For the whole human genome we used the same fragment
size as for the horse genome. The match and mismatch costs are 
optimized for accuracy on the 251bp length {\it E.coli} reads. For other genomes we recommend
the user to experiment with different fragment sizes starting with a small value. As explained
earlier the MaxSSmap fragment length must be a multiple of 16 because
of byte packing to allow storage of the genome in an array of int4 instead of char.

\paragraph{MaxSSmap\_fast}
In this faster version of MaxSSmap we consider every other nucleotides in 
the read sequence when mapping to the genome. This heuristic reduces runtime 
considerably than if we were to compare all nucleotides in the read sequence.
See files MaxSSMap\_shared\_int4\_fast.cu 
in the source code for exact implementation.

\paragraph{SSW}
This is a recent Smith-Waterman library that uses  Single-Instruction-Multiple-Data (SIMD)
to achieve parallelism. It has been shown to be faster than other SIMD based Smith-Waterman
approaches \cite{ssw}. It has also been applied to real data as a
secondary program to align reads rejected by primary programs \cite{ssw}.

\paragraph{CUDA-SW++}
CUDA-SW++ \cite{cudasw++} is originally designed for
protein database search. It performs Smith-Waterman alignment of the
query to each sequence in the database in parallel. We simulate short read mapping with it
by dividing the genome into same size disjoint fragments and 
considering each fragment of the genome as one of the database sequences
and the read as the query. We set CUDA-SW++ to output the two top highest scoring fragments 
and their scores. If the ratio of the second 
best score to the best one is above .9 we do not consider the read mapped.
We set the fragment length to 512 and 2400 for the {\it E.coli} genome and horse genomes,
the gap open and extension costs to -26 and -1, and the match and 
mismatch costs to 5 and -4. These values yielded
highest accuracy for the simulated reads. We modified the code so that the 
blosum45 matrix uses +5 for match and -4 for mismatch.
We choose 512 fragment length for {\it E.coli} because lower ones reduce the runtime marginally 
but the accuracy goes down considerably whereas higher fragment lengths don't 
yield higher accuracy and increase runtime. The gap, match, and mismatch costs are optimized for 
accuracy on the 251bp {\it E.coli} reads. For the horse genome we couldn't run CUDA-SW++
with higher fragment lengths of 4800 and so we selected 2400. 

\paragraph{BWA-MEM}
We use BWA-MEM version 0.7.5a with multi-threaded enabled (-t 12) and other options
set to their default values.

\paragraph{NextGeneMap}
We use NextGeneMap version 0.4.10 with the options -g 0 that enables the GPU
 and everything else default.

\paragraph{Meta-methods} We consider four meta-methods that first apply a 
NextGenMap and then a more accurate aligner for rejected reads.
\begin{itemize}
\itemsep -1pt
\item NextGenMap + MaxSSmap
\item NextGenMap + MaxSSmap\_fast
\item NextGenMap + CUDASW++
\item NextGenMap + SSW
\end{itemize}

We use the same options for each program in the meta-method as described above.

\subsection{Experimental platform}
All programs were executed on Intel Xeon X5650 machines with
12GB RAM each equipped with three NVIDIA Tesla M2050 GPUs 
with 3GB global memory and 49152 byes of shared memory. 
We used CUDA release 4.2 to develop MaxSSmap and to compile and
build the GPU programs. In Table~\ref{programplatform} we list the architecture on 
which we run each program.

\subsection{Data simulation}
We use the program Stampy \cite{stampy} (version 1.0.22) 
to simulate reads with realistic base qualities. We use the {\it E.coli} genome 
K12 MG1665 (4.6 million bp) from which Stampy simulates reads
and Illumina MiSeq 251bp reads in SRR522163 from the NCBI Sequence Read Archive 
(\url{http://www.ncbi.nlm.nih.gov/Traces/sra}) from which Stampy simulates base qualities. 
For the human reads we use the human chromosome one sequence from the 
Genome Reference Consortium 
{(\url{http://www.ncbi.nlm.nih.gov/projects/genome/assembly/grc/)}}
version GRCh37.p13. We use Illumina MiSeq 250bp reads in ERR315985 through ERR315997  
from the the NCBI Sequence Read Archive from which Stampy simulated base qualities. 

We simulate one million 251 bp {\it E.coli} reads and 250bp human chromosome one
reads of divergences 0.1, 0.2, and 0.3  with and without gaps ranging upto length 30. 
The gaps are randomly chosen to occur in the read or the genome.
Roughly speaking each divergence corresponds to fraction of mismatches in the
reads after accounting for sequencing error. For example .1 divergence means on 
average 10\% mismatches excluding sequencing errors.
See Table~\ref{stampysimulation} for exact Stampy command line parameters for simulating
the data.

\subsection{Measure of accuracy and error}
For Stampy simulated reads the true alignment is given in a CIGAR string format \cite{sam}.
Except  for CUDA-SW++ we evaluate the accuracy of all programs with the
same method used in \cite{stampy}. We consider the read to be aligned
correctly if at least one of the nucleotides in the read is aligned to the 
same one in the genome as given by the true alignment. It's not unusual to
allow a small window of error as done in previous studies (see \cite{ayat}
for a thorough discussion). 

CUDA-SW++ does not output in SAM format. Instead it gives the top scoring fragments
and the score of the query against the fragment. To evaluate 
its accuracy we divide the true position by the fragment size 
which is 512 for {\it E.coli} and 2400 for horse genome in our experiments. 
We then consider the read to be mapped correctly if the 
difference between the CUDA-SW++ fragment and the true one is at most 1. 

\section{Results}
We study the accuracy and runtime of all programs and the four meta-methods
described earlier. We measure their performance for 
mapping simulated Illumina {\it E.coli} and human reads to the {\it E.coli} and human
chromosome one respectively. We then compare them on reads of different lengths
and demonstrate applications on real data.

\subsection{Comparison of MaxSSmap and Smith-Waterman for mapping 
divergent reads to {\it E.coli} genome}
We begin by comparing MaxSSmap and MaxSSmap\_fast to SSW and CUDA-SW++.
We map 100,000 251bp simulated {\it E.coli} reads
to the {\it E.coli} genome. We simulate these reads using the Stampy \cite{stampy} 
program (described earlier).

As mentioned earlier, MaxSSmap offers no advantage over hash-table approaches 
on reads with fewer than 10\% mismatches. Programs like NextGenMap 
\cite{nextgenmap} designed for mapping
to polymorphic genomes can align such reads very quickly with high accuracy. 
Thus we consider three levels of divergence in the reads: 0.1, 
0.2, and 0.3. Roughly speaking each divergence corresponds to the percentage
of mismatches in the data. 

In Table~\ref{ecoli2}(a) we see that the
MaxSSmap accuracy is comparable to SSW and CUDA-SW++ except
at divergence 0.3 with gaps (our hardest setting). Table~\ref{ecoli2}(b) shows 
that the MaxSSmap and MaxSSmap\_fast
runtimes are at least 44 and 60 times lower than SSW and 5.8 and 8 times lower
than CUDA-SW++. This is where the real advantage of MaxSSmap lies:
high accuracy and low error comparable to Smith-Waterman on reads up to
30\%  mismatches and gaps yet at a lower cost of runtime.

At high divergence and with gaps we expect Smith-Waterman
to fare better in accuracy and error than our maximum scoring subsequence heuristic.
For example at divergence 0.3 with gaps SSW is 5.1\% and
14\% better than MaxSSmap and MaxSSmap\_fast in accuracy.

Recall that MaxSSmap  detects and rejects repeats which are likely to be errors.
We use the same technique in the CUDA-SW++ output. However,
SSW does not appear to have such a strategy and so we see a higher error for it.

\subsection{Comparison of meta-methods for mapping divergent {\it E.coli} reads}
We now compare the accuracy and runtime of  four meta-methods
that use NextGenMap in the first phase of mapping and MaxSSmap, MaxSSmap\_fast,
CUDASW++, and SSW to align rejected reads in the second phase. 
We study the mapping of one million 251 bp reads simulated {\it E.coli} reads 
to the {\it E.coli} genome. 

In Table~\ref{ecoli3} we see that the accuracy of NGM+CUDASW++ 
and NGM+SSW are comparable to NGM+MaxSSmap 
but runtimes are much higher. For example at divergence 0.2 with gaps
NGM+SSW takes over 48 hours to finish and NGM+CUDASW++ takes 756 minutes, whereas NGM+MaxSSmap and NGM+MaxSSmap\_fast finish in 109 and 68 minutes 
respectively. At divergence 0.3 with gaps both NGM+MaxSSmap and NGM+MaxSSmap\_fast
finish within four hours whereas both NGM+CUDASW++ and NGM+SSW take more than
24 hours. We choose the two fastest meta-methods for comparison
to BWA and NextGenMap. 

\subsection{Comparison of fastest meta-methods to NextGenMap and BWA for mapping  
divergent {\it E.coli} and human chromosome one reads}
In Tables~\ref{ecoli} and ~\ref{humanchr1} we compare the accuracy 
and runtimes of NextGenMap and BWA to 
NGM+MaxSSmap and NGM+MaxSSmap\_fast. Both meta-methods achieve high
accuracy and low error at all settings but at the cost of increased runtime compared
to NextGenMap and BWA. On {\it E.coli} reads of divergence 0.1 and 0.2 with gaps 
NextGenMap+MaxSSmap\_fast yields an improvement of 14\% and 32\% over NextGenMap
while adding 37 and 66 minutes to the NextGenMap time of 1.5 and 2 minutes 
respectively. On human chromosome 1 reads of the same settings 
NextGenMap+MaxSSmap\_fast correctly aligns an additional 2\% and 
13\% reads than NextGenMap alone at the cost of 3.2 and 582 extra minutes. 
The MaxSSmap runtimes for human chromosome 1 are higher than for 
{\it E.coli} because there are many more fragments to consider in the former.
The runtimes for both meta-methods also
increases with higher divergence because there are many more reads rejected by 
NextGenMap at those divergences.

NextGenMap has higher accuracy than BWA as shown here in Tables~\ref{ecoli} and~\ref{humanchr1}
and in previous studies \cite{nextgenmap} while BWA is the fastest program amongst
all compared. We ran BWA in a multi-threaded mode that utilizes all CPU cores and all 
other methods on the GPU. We found that running NextGenMap on the GPU was faster
than its multi-threaded mode.

\subsection{Comparison of methods on reads of various lengths}
The simulated sequences in our above results are based upon Illumina MiSeq sequences
of lengths 250 and 251 bp. Here we study simulated reads of lengths 36, 51, 76, 100,
and 150 based on real E.coli sequences from the Illumina Genome Analyzer II (36bp), 
HiSeq 1000 (51bp), HiSeq 2000 (76 and 100bp), and MiSeq (150bp). We obtained 
the sequences from datasets ERR019652 (36bp), SRR1016504 (51bp), SRR1016920 (76bp),
ERR376625 (100bp), and SRR826444 and SRR826446 (150bp) 
in the NCBI Sequence Read Archive (\url{http://www.ncbi.nlm.nih.gov/Traces/sra}). 
We simulated 1 million reads of each length from the {\it E.coli} genome and of divergence 
0.1 with gaps (up to length 30) with Stampy as described earlier. Recall that Stampy 
simulated base qualities based upon the real data in the input.

In Table~\ref{readlengths} we compare the accuracy and runtimes of BWA, NextGenMap,
NextGenMap+MaxSSmap\_fast, and NextGenMap+MaxSSmap. As 
the read length increases we see that NextGenMap and
the meta-methods increase in accuracy and decrease in error. BWA is the most 
conservative and has lowest error at all lengths especially on the shortest read lengths. 
At reads lengths of 100 and above
NextGenMap+MaxSSmap has about 10\% higher accuracy than NextGenMap and 1\%
more error. 

\subsection{Comparison to parallel multi-threaded CPU implementation of MaxSSmap}
We also study the runtimes of the parallel multi-threaded CPU implementation of 
MaxSSmap as described earlier. We examined three fragments lengths of 4800,
48000, and 480000. Each yields 1042, 104, and 11 threads to run on available CPU
cores. We ran this program on Intel Xeon CPU which has a total of 12 cores.

We tested this for mapping a 100,000 251 bp {\it E.coli} reads and found fragment length of
4800 to be the fastest. We then mapped 100,000 251 bp {\it E.coli} reads which took
224 minutes. In comparison the GPU MaxSSmap takes 20 minutes. Thus we find the
multi-threaded version to be 10 times slower.

\subsection{Applications on real data}
We consider two scenarios in real data where unmapped divergent reads may occur. 
The first is in the mapping of ancient fossil DNA to modern genomes and second is the
alignment of unmapped reads when comparing a human genome to the standard reference.

\subsubsection{Ancient horse DNA mapping to modern genomes}
For the first case we consider reads obtained from an ancient horse bone
\cite{ancienthorseDNA}. In a previous study the parameters of the BWA program were
optimized to maximize mapped reads from this set to the horse and human genomes
\cite{ancienthorse}. We consider one set of reads from the same study in dataset 
SRR111892 obtained from the 
NIH Sequence Read Archive (\url{http://www.ncbi.nlm.nih.gov/Traces/sra}). 
These reads are produced by the
Illumina Genome Analyzer II sequencer and have an average length of 67.7
and standard deviation of 8.4 
We obtained the human genome from the Genome Reference Consortium 
{(\url{http://www.ncbi.nlm.nih.gov/projects/genome/assembly/grc/)}}
version GRCh37.p13 and the horse genome
Equus\_caballus EquCab2 (GCA\_000002305.1) from Ensemble (\url{http://useast.ensembl.org/Equus\_caballus/Info/Index}).

Highly divergent sequences are likely to be present in this dataset and as we
have seen in the previous section short read lengths of up to 76  
are challenging even with 10\% mismatches and gaps. Thus 
we consider only reads of the maximum
length of 76 in this dataset. We map the first 100000 
such reads with BWA, NextGenMap,
NextGenMap+MaxSSmap\_fast, NextGenMap+MaxSSmap, and
NextGenMap+CUDASW++ to the horse and human genomes. 
We consider the human genomes to identify ancient 
human DNA fragments in the sample \cite{ancientDNA}. 

In Table~\ref{ancienthorse} we see that NextGenMap aligns 16\% and 14\% of the 
reads to the horse and human genomes whereas NextGenMap+MaxSSmap aligns
a total of 23.1\% and 21\% respectively. BWA in comparison aligns 2.2\% and 0.16\% with
default parameters and has similar accuracy with optimized parameters given 
in a previous study \cite{ancienthorse}. 

We evaluated NextGenMap+CUDA-SW++ for mapping reads to the horse genome 
by running it for a maximum of 168 hours (one week). In 
that time period CUDA-SW++ considered 56800
of the 84036 NextGenMap rejected reads to be aligned  (about 68\%) and mapped 
11.8\% of them. Based on this we estimate it would 
take 247 hours for CUDA-SW++ to consider all of the NextGenMap rejected reads. 
And it would align a total of 11.8\% of 84036 reads which equals 9916. This
added to the NextGenMap mapped reads gives a total mapping rate of 26\%.

To better understand these numbers we simulated 100,000 horse 76 bp reads 
of divergence 40\% and gaps with Stampy and with base qualities simulated 
from the same real dataset  (SRR111892) used here. These settings 
were chosen to achieve a ballpark mapping
rate with the real data. We find that NextGenMap aligns 18\% of total reads with only 0.4\%
correctly mapped while BWA maps no reads at all. These are similar to
the mapping rates on the real data. This suggests these are difficult settings 
for NextGenMap and BWA.
When we apply MaxSSmap to reads missed by NextGenMap it aligns an additional 
8.8\% (similar to the real data) with 3.3\% correct. To reduce the error to zero 
we raise the MaxSSmap mapping
quality threshold from the default 15.2 to 62. Raising it gives fewer but higher
quality mapped reads. This gives 0.112\% mapped reads all of which are correct.

With this in mind we return to the real data and apply NextGenMap+MaxSSmap with the higher
mapping quality cutoff (of 62) in the hopes of obtaining all correct alignments. 
The higher threshold gives 9 additional reads. In the Supplementary Material we 
give the SAM output of these alignments. Most have at least 30\% mismatches and
challenging for both BWA and NextGenMap.

\subsubsection{Mapping unaligned reads from NA12878 to human reference}
For our second scenario we consider the popular human genome sequence 
NA12878 and study the mapping of one of its dataset SRR16607 to the 
same human reference used above. We map the first 100000 paired 
reads in SRR016607 (101bp) with NextGenMap,
NextGenMap+MaxSSmap\_fast, and 
NextGenMap+MaxSSmap (fragment length set to 4800). In the latter two
methods we re-align the pairs with MaxSSmap\_fast and MaxSSmap where at least one read in 
the pair was unmapped by NextGenMap or the mapped pair positions were
outside the mapping distance threshold of 500 base pairs. 

In Table~\ref{SRR016607} we measure the number
of paired reads whose mapped positions are within 500 base pairs. 
Although the nominal insert size of this dataset is 300 with a standard deviation of
77 (as given in the NIH SRA website) we found many mapped pairs in the output that
were within 500 base pairs and so we use this threshold. These mapped pairs are 
called concordant reads \cite{stampy}. We see that NextGenMap aligns 83.5\% of the reads
concordantly whereas NextGenMap+MaxSSmap\_fast and NextGenMap+MaxSSmap 
align 85.5\% and 87.3\%. Both methods mapped 0.7\% and 1.2\% pairs discordantly
(mapped positions at least 500 bp apart). NextGenMap mapped 9.4\% of the pairs 
discordantly but we don't report this here because we re-align those reads with MaxSSmap.


\section{Discussion}
In our experimental results we have demonstrated the advantage of MaxSSmap 
over Smith-Waterman for mapping reads to genomes. In scenarios where accurate
re-alignment of rejected and low-scoring reads are required MaxSSmap and MaxSSmap\_fast
would be fast alternatives to Smith-Waterman. Such conditions are likely to contain
reads with many mismatches and gaps which would get rejected by 
programs based on hash-tables and Burrows-Wheeler. 

We demonstrate two such scenarios on real data. In both cases we see an
increase in mapped reads by MaxSSmap. While this increase 
comes at the cost of considerable runtime it is still much
faster than the Smith-Waterman alternative. Furthermore, the output of
NextGenMap+MaxSSmap reveals many high scoring alignments
well above the mapping quality threshold that warrant further study.
The MaxSSmap alignments of horse DNA reads to the horse and human genomes
contain 39.3\% and 39.4\% mismatches on the average. In the human genome
paired read study we find the MaxSSmap concordant aligned pairs to contain
19.1\% mismatches on the average. In both cases these are challenging divergences
for BWA and NextGenMap as we saw in the simulation studies. 

Our real data applications in this paper are brief and deserve a wider study in
a separate paper. For example when we aligned unmapped pairs in NA12878 
to the human genome we obtained 3.8\% more pairs with MaxSSmap.
We will search for variants in these alignments as part of future work. Also
in future work we plan to study MaxSSmap on metagenomic reads where
divergence rates can be high as well.

Our results on both real and simulated data are an insight into missed reads
that are rejected by BWA and NextGenMap. We see that the high mismatch rate
and gaps are the main reasons why these reads are unmapped in the first
place. Without more exact approaches like MaxSSmap and Smith-Waterman it
would be much harder to align such reads.

The ratio of the NextGenMap+CUDASW++ to the NextGenMap+MaxSSmap runtimes 
varies and can depend upon number of reads to align. In Table~\ref{ecoli3} where
reads lengths are fixed at 251bp we see that this ratio is 5.8 at divergence 0.1
without gaps where both MaxSSmap and CUDA-SW++ align approximately
12,000 reads (that number rejected by NextGenMap in Table~\ref{ecoli}). 
As the divergence increase to 0.3 with gaps both programs align about 
800,000 reads and there the ratio of their runtimes is 6.9.

Our program is not without limitations. We find that at very short reads 
of lengths 36 and 51 the improvements given by MaxSSmap are small
compared to higher lengths. 
However, read lengths of 125 and above are not uncommon 
especially since current Illumina machines such as MiSeq, HiSeq, and NextSeq
generate reads of at least this length (see \url{http://www.illumina.com/systems/sequencing.ilmn})

We find the runtimes are much higher for human
and horse chromosomes than for {\it E.coli} just because there are many more genome 
fragments for the former. This could be lowered by spreading reads across multiple
GPUs. We also see that the accuracy of MaxSSmap is lower than 
Smith-Waterman as we cross into higher divergence of 0.3 with gaps.

\section{Conclusion}
We introduce a GPU program called MaxSSmap for mapping reads to genomes. We
use the maximum scoring subsequence to identify candidate genome fragments for
final alignment instead of hash-tables and Burrows-Wheeler transform. We show
that MaxSSmap has comparable high accuracy to Smith-Waterman based programs yet
has lower runtimes and accurately maps reads rejected by a hash-table 
based mapper faster than if Smith-Waterman were used.
We also study MaxSSmap on different read lengths and demonstrate
applications on real data by mapping ancient horse DNA reads to modern genomes
and unmapped paired reads from NA12878 in 1000 genomes.
\bigskip

\section{Competing interests}
The authors declare that they have no competing interests.

\section{Author's contributions}
U.R. designed and implemented MaxSSmap. T.T. and U.R. conducted
all experiments. 

\section{Acknowledgements}
We thank Peicheng Du, Jeffrey Rosenfeld, and anonymous reviewers for 
helpful suggestions to our paper. This research was performed on a GPU cluster at 
UNC Charlotte thanks to Dennis R. Livesay 
and a GPU machine at NJIT thanks to
Shahriar Afkhami. Funding was provided by the National
Science Foundation Assembling Tree of Life grant 0733029.


\bibliographystyle{unsrt} 
\bibliography{my_bib}      




\section{Figures}

\section{Tables}

\begin{table}[h]
\small
\centering
\caption{Architecture for each program compared in our study \label{programplatform}}
\begin{tabular}{ @{} l  l @{} } \hline 
Program   &Architecture \\ \hline
MaxSSmap & GPU \\ 
MaxSSmap\_fast & GPU \\
CUDA-SW++ & GPU \\
SSW & SIMD single CPU (GPU unavailable)\\
NextGenMap & GPU \\
BWA-MEM & multi-threaded 12 CPUs \\
\end{tabular}
\end{table}

\begin{table}[h]
\small
\centering
\caption{Stampy (version 1.0.22) parameters to simulate reads \label{stampysimulation} }
\begin{tabular}{@{} l  l @{}} \hline 
Genome   & Stampy  parameters  \\ \hline
{\it E.coli} (format genome)  & -G  ecoli ecoli\_K12\_MG1665.fasta \\
(hash)           & -g  ecoli -H ecoli  \\
(simulate)           & -g  ecoli -h ecoli -S SRR522163\_1.fastq \\ \hline
Human (format genome)& -G hs\_ref\_GRCh37.p13\_chr1 \\
             &   hs\_ref\_GRCh37.p13\_chr1.fa \\
(hash)    & -g hs\_ref\_GRCh37.p13\_chr1 \\
	   & -H hs\_ref\_GRCh37.p13\_chr1 \\
(simulate)         & -g hs\_ref\_GRCh37.p13\_chr1 \\
		& -h hs\_ref\_GRCh37.p13\_chr1 \\
	&  -S ERR315985\_to\_ERR315997\_1.fastq \\ \hline
	Divergence   &     \\ 
.1    & --substitutionrate=.1 \\
.2    & --substitutionrate=.2 \\
.3    & --substitutionrate=.3 \\ 
.1+gaps & --substitutionrate=.1 --simulate-minindellen=-30\\ 
              & --simulate-maxindellen=30 --insertsize=250 \\
              & --insertsd=25 \\
.2+gaps & --substitutionrate=.2 --simulate-minindellen=-30 \\
		& --simulate-maxindellen=30 --insertsize=250 \\
		& --insertsd=25 \\
.3+gaps & --substitutionrate=.3 --simulate-minindellen=-30\\
		&  --simulate-maxindellen=30 --insertsize=250 \\
		& --insertsd=25 \\ \hline
\end{tabular}
\end{table}

\begin{table}[h]
\small
\centering
\caption{Comparison of MaxSSmap and MaxSSmap\_fast to a GPU and
a SIMD high performance Smith-Waterman implementation. 
These are simulated Illumina reads and contain
realistic base qualities generated from Illumina short reads. Each divergence
represents the average percent of mismatches in the reads. So 0.1 means 
10\% mismatches on the average. The gaps are randomly chosen to
occur in the read or the genome and are of length at most 30. \label{ecoli2} }
\subfloat[Percent of 100,000 251bp reads mapped correctly to
the {\it E.coli} genome.  Shown in parenthesis are 
incorrectly mapped reads and remaining are rejected. ]{
\begin{tabular}{ @{} l  l  l  l  l @{} } \hline 
Div   & MaxSSmap\_fast & MaxSSmap & CUDASW++ & SSW \\ \hline
\multicolumn{5}{c}{Reads without gaps} \\
.1    &   95 (0.4)  & 96 (0.4) & 94 (0.9)  &  97 (3)\\
.2    &   95 (0.6)  & 95.3 (0.6) & 94 (1)     &  97 (3) \\
.3    &   90 (1.1) & 94.2 (0.9) & 93 (1.3) &   96 (4) \\ \hline
\multicolumn{5}{c}{Reads with gaps} \\
.1    &  92 (1.5) &  93.1 (1.9)   &  94 (0.9) &  97 (3)  \\
.2    & 90 (1.7) &   92.5 (2.1)    & 92 (1)    &   96 (4)  \\
.3    & 81 (2.8)  &  89.9 (3.5)   &  92 (1.4) &  95 (5)    \\ \hline
\end{tabular}
}
\qquad
\subfloat[Time in minutes to map 100,000 251bp reads to
the {\it E.coli} genome.]{
\begin{tabular}{ @{} l  l  l  l  l @{} } \hline 
Div   & MaxSSmap\_fast  & MaxSSmap & CUDASW++ & SSW \\ \hline
\multicolumn{5}{c}{Reads without gaps} \\
.1    & 20  & 28 & 164 & 1288 \\
.2    & 20  & 28 & 164 & 1275  \\
.3    & 20  & 28 & 164 & 1255 \\ \hline
\multicolumn{5}{c}{Reads with gaps} \\
.1    & 20  & 28 & 163 &    1283     \\
.2    & 20  &  28 & 162 &    1266    \\
.3    & 20   & 28 & 162 &    1235    \\ \hline
\end{tabular}
}

\end{table}

\begin{table}[h]
\small
\centering
\caption{Comparison of meta-methods. See Table~\ref{ecoli2} caption for details about reads.
NA denotes time greater than 48 hours which is our cutoff time on this data. \label{ecoli3}}
\subfloat[Percent of one million 251bp reads mapped correctly to
the {\it E.coli} genome. Shown in parenthesis are incorrectly mapped reads and
remaining are rejected.]{
\begin{tabular}{  @{} l  l  l  l  l @{} } \hline 
Div   & NextGenMap+  & NextGenMap+ &  NextGenMap+   & NextGenMap+ \\  
       &  MaxSSmap\_fast &  MaxSSmap &  CUDASW++   & SSW \\ \hline
\multicolumn{5}{c}{Reads without gaps} \\
.1     &  96 (1.3) & 97 (1.4)    & 97 (1)   & 97 (2.8) \\
.2    &  95 (1.5)  & 96.7 (1.5)   &   96 (1)   & NA \\
.3    & 91  (1.4) &  94.8 (1.3)    &  93 (1.5)   &  NA \\ \hline
\multicolumn{5}{c}{Reads with gaps} \\
.1      & 92 (1.5) & 95.7 (1.4)    &   97 (1)  &   97.2 (2.8) \\
.2      & 92 (2.1) &  94  (2.5)   & 95 (2) &  NA  \\
.3    &  82 (2.9) & 90.5 (3.5)  &  92.5  (1.6)  &   NA \\ \hline
\end{tabular}
}
\quad\quad
\subfloat[Time in minutes to map one million 251bp reads to
the {\it E.coli} genome.]{
\begin{tabular}{ @{} l  l  l  l  l @{} } \hline 
Div   & NextGenMap+  & NextGenMap+ &  NextGenMap+   & NextGenMap+ \\  
       &  MaxSSmap\_fast &  MaxSSmap &  CUDASW++   & SSW \\ \hline
\multicolumn{5}{c}{Reads without gaps} \\
.1    & 25 & 34       &  197  &  1397   \\
.2    &  57  &  77     &  537   &  NA   \\
.3    & 148  & 204   &  1343 & NA    \\ \hline
\multicolumn{5}{c}{Reads with gaps} \\
.1    & 38  &  60   &  413  &  2601  \\
.2    & 68 &   109 &   756   & NA  \\
.3    &  162 & 222  &  1528 &  NA  \\ \hline
\end{tabular}
}
\end{table}

\begin{table}[h]
\small
\centering
\caption{Comparison of meta-methods to NextGenMap and BWA. 
See Table~\ref{ecoli2} caption for details about reads.\label{ecoli}}
\subfloat[Percent of one million 251bp reads mapped correctly to
the {\it E.coli} genome. Shown in parenthesis are incorrectly mapped reads and
remaining are rejected.]{
\begin{tabular}{  @{} l  l  l  l  l @{} } \hline 
Div   & BWA  & NextGenMap & NextGenMap+  & NextGenMap+ \\  
       &           &                     & MaxSSmap\_fast &  MaxSSmap \\ \hline
\multicolumn{5}{c}{Reads without gaps} \\
.1     &  89 (1.1)   &  87 (1) &  96 (1.3) & 97 (1.4)     \\
.2    &  24 (0.5) &  72 (1)   &  95 (1.5)  & 96.7 (1.5)  \\
.3    &  0.6 (0)   &  26 (0.5)&  91  (1.4) &  94.8 (1.3)  \\ \hline
\multicolumn{5}{c}{Reads with gaps} \\
.1      & 85 (3)  &  78 (1)   &  92 (1.5) & 95.7 (1.4)      \\
.2      & 20 (0.5) & 60 (1)    & 92 (2.1) &  94 (2.5)    \\
.3    &  0.5 (0)    &  19 (0.4) &  82 (2.9)  &  90.5 (3.5)  \\ \hline
\end{tabular}
}
\quad\quad
\subfloat[Time in minutes to map one million 251bp reads to
the {\it E.coli} genome.]{
\begin{tabular}{ @{} l  l  l  l  l @{} } \hline 
Div   & BWA  & NextGenMap & NextGenMap+  & NextGenMap+ \\  
       &           &                     & MaxSSmap\_fast &  MaxSSmap \\ \hline
\multicolumn{5}{c}{Reads without gaps} \\
.1    & 0.7  &  1.2   &  25 &  34        \\
.2    & 0.5  &  1.9    &  57  &  77      \\
.3    & 0.4  &  2.1    &  148  & 204    \\ \hline
\multicolumn{5}{c}{Reads with gaps} \\
.1    & 0.7  &  1.5  &  38  &  60    \\
.2    & 0.5  &  2.1  & 68 &   109    \\
.3    & 0.4  &  2.1   &  162 &  222 \\ \hline
\end{tabular}
}
\end{table}

\begin{table}[h!]
\small
\centering
\caption{Comparison of meta-methods to NextGenMap and BWA. 
See Table~\ref{ecoli2} caption for details about reads. \label{humanchr1} }
\subfloat[Percent of one million 250bp reads mapped correctly to
the human chromosome one genome. Shown in parenthesis are incorrectly mapped reads and
remaining are rejected.]{
\begin{tabular}{  @{} l  l  l  l  @{}  } \hline 
Div   & BWA  & NextGenMap & NextGenMap+   \\  
       &           &                     & MaxSSmap\_fast \\ \hline
\multicolumn{4}{c}{Reads without gaps} \\
.1   & 96(3)    &   99 (1)    &  99 (1)      \\
.2   & 33 (6)   &  86 (6)     &  94 (6.2) \\
.3   &  0.8 (6)  & 37 (9)      &   88 (10.5)    \\ \hline
\multicolumn{4}{c}{Reads with gaps} \\
.1  &  89 (3)     &  96 (2)    &   98 (2)      \\
.2   &  28 (5)  &   79 (6)    &   92 (6.5)   \\
.3   &  0.7 (0.5) &  30 (7)   &    80 (9.4)  \\ \hline
\end{tabular}
}
\qquad
\subfloat[Time in minutes to map one million 250bp reads to
the human chromosome one genome.]{
\begin{tabular}{ @{} l  l  l  l  @{}  } \hline 
Div   & BWA  & NextGenMap & NextGenMap+   \\  
       &           &                     & MaxSSmap\_fast  \\ \hline
\multicolumn{4}{c}{Reads without gaps} \\
.1  &  1.6  &  7.1  &  10.3     \\
.2  &  0.8  &  44   &  626     \\
.3  &   0.6  &  65   &  3992     \\ \hline
\multicolumn{4}{c}{Reads with gaps} \\
.1 & 1.5  &  9.8   & 176    \\
.2 & 0.7  &  51  &  1126   \\
.3  &  0.6  & 66  & 4611    \\ \hline
\end{tabular}
}
\end{table}


\begin{table}[h!]
\small
\centering
\caption{Percent of one million reads of lengths 36, 51, 76, 100, and 150 and of
divergence 0.1 with gaps mapped correctly to
the {\it E.coli} genome. Shown in parenthesis are incorrectly mapped reads and
remaining are rejected. We denote NextGenMap by NGM. \label{readlengths} }
\begin{tabular}{  @{} l  l  l  l  l  l @{}  } \hline 
Read  & BWA & NGM & NGM+           & NGM+  & NGM+ \\ 
length                   &          &       & MaxSSmap\_fast & MaxSSmap         & CUDA-SW++\\
                    &        &        &                        &     \\ \hline  
36         &    1.2 (0)    &     33.3 (10.5)   &   34 (16.5)      &  43.8 (18.1) & 53.2 (17.1) \\ 
51          &    11.4 (0.2) &    50.5 (3)         &  53.3 (8.8)   &  71.2 (7.5)  &  79.5 (6.2) \\
76         &   38.4 (.7)  &    70.2 (1.5)        &   76.5 (4.3)   &  92.9 (2.7) & 96.6 (1.8) \\
100        &   62.4 (1)     & 82.6 (1.4)    &   91 (2.2)       &  96.8 (1.9)  &  98.1 (1.5) \\
150       &     76 (1.1)      & 82.5 (1.2)     &   92 (2.2)    &  93.8 (2.2)  & 94.7 (2.1)  \\ \hline
\multicolumn{6}{c}{Time in minutes to map reads} \\ 
36       &     0.09    &  0.3   & 26.6     & 37.8  & 315.9 \\       
51     &     0.11      & 0.5   &  26.6      & 38.2 & 284.9 \\
76      &   0.15      & 0.6    &  19.8   & 28.4  & 201.9 \\
100     &   0.22      & 0.7   &  14.8    &  21  &  129.9  \\ 
150      &    0.34     &  0.9    &  19.6 &  29.2  & 173.8  \\ \hline
\end{tabular}
\end{table}

\begin{table}[h!]
\small
\centering
\caption{Percent of 100,000 ancient horse DNA reads (SRR111892) of length
76 bp mapped to the horse genome Equus\_caballus 
EquCab2 (GCA\_000002305.1) and human reference genome. 
We ran NextGenMap+CUDA-SW++ for a maximum of 168 hours and estimated
the time to align all rejected reads. Also shown is time in minutes. \label{ancienthorse} }
\begin{tabular}{  @{} l  l  l  l  l @{}  } \hline 
\multicolumn{5}{c}{Horse genome} \\
BWA & NextGenMap &   NextGenMap+ & NextGenMap+  & NextGenMap+\\
       &                         &   MaxSSmap\_fast & MaxSSmap  &  CUDASW++\\
 2.2 &  16              &     20.5         &  23.1         &    26 (estimated) \\ \hline
\multicolumn{5}{c}{Time in minutes to map reads}  \\
 0.6 &      2.4         &  1609.6       &      2836    &   14820 (estimated)\\ \hline \hline
\multicolumn{5}{c}{Human genome} \\
BWA & NextGenMap &   NextGenMap+ & NextGenMap+  & NextGenMap+ \\
       &                         &   MaxSSmap\_fast & MaxSSmap  &  CUDASW++ \\
  0.16 &  14              &   18.6         &  21   &   NA   \\ \hline
\multicolumn{5}{c}{Time in minutes to map reads}  \\
   0.82   &  2.9 &  2375.5     &    4108     &       NA   \\      
\end{tabular}
\end{table}

\begin{table}[h!]
\small
\centering
\caption{Percent of 100,000 paired human reads from NA12878 in 1000 genomes 
(SRR016607) of length 101 bp mapped concordantly to the human genome. 
Concordant reads are pairs that are mapped within 500 base pairs.
Also shown in parenthesis are discordant reads (mapped positions 
at least 500 bp apart) and the time in minutes.
In NextGenMap+MaxSSmap we re-align pairs with MaxSSmap where at least one read in 
the pair was unmapped by NextGenMap or the pair is discordant.
Thus, NextGenMap shows zero discordant pairs because we re-align them with
MaxSSmap. \label{SRR016607} }
\begin{tabular}{  @{} l  l l @{}  } \hline 
NextGenMap &  NextGenMap+        & NextGenMap+ \\  
                     & MaxSSmap\_fast     & MaxSSmap  \\ 
       83.5 (0)  &     85.5 (0.7)   &  87.3 (1.2) \\  \hline
\multicolumn{3}{c}{Time in minutes to map paired reads} \\ 
         1.5        &    1256.1      &       2199.1   \\       
\end{tabular}
\end{table}

\end{document}